\documentclass[aps,nofootinbib,floatfix,prd,letterpaper,twocolumn]{revtex4} 
\usepackage{graphicx,amsmath,amssymb,color,bm}
\usepackage{hyperref}
\def\be{\begin{equation}}
\def\ee{\end{equation}}
\def\bea{\begin{eqnarray}}
\def\eea{\end{eqnarray}}
\newcommand{\vs}{\nonumber\\}
\def\ba#1\ea{\begin{align}#1\end{align}}
\def\bg#1\eg{\begin{gather}#1\end{gather}}
\def\Msunh{h^{-1}M_{\odot}}
\def\iMpch{\,h\,{\rm Mpc}^{-1}}
\def\cH{\mathcal{H}}

\def\rhob{\bar{\rho}}
\def\O{\mathcal{O}}

\newcommand{\s}{\sigma}
\newcommand{\refeq}[1]{Eq.~(\ref{eq:#1})}          
\newcommand{\refeqs}[2]{Eqs.~(\ref{eq:#1})--(\ref{eq:#2})}          
\newcommand{\reffig}[1]{Fig.~\ref{fig:#1}}          
\newcommand{\refsec}[1]{Sec.~\ref{sec:#1}}          
\newcommand{\refapp}[1]{App.~\ref{app:#1}}

\renewcommand{\v}[1]{\mathbf{#1}}
\newcommand{\vx}{\v{x}}
\renewcommand{\vr}{\v{r}}
\newcommand{\vy}{\v{y}}

\newcommand{\vk}{\v{k}}
\newcommand{\vq}{\v{q}}

\newcommand{\<}{\langle}
\renewcommand{\>}{\rangle}

\renewcommand{\d}{\delta}
\newcommand{\eps}{\epsilon}
\newcommand{\Om}{\Omega_m}

\def\olp{\text{1-loop}}

\def\intMs{\int_{\ln M_s}\!\!\!\!}

\def\Mbar{\< M \>_{\rho}}

\begin{document}

\title{Towards a self-consistent halo model for the nonlinear large-scale structure}

\author{Fabian Schmidt}
\affiliation{Max-Planck-Institut f\"ur Astrophysik, Karl-Schwarzschild-Str.~1, 85748~Garching, Germany}

\begin{abstract}
The halo model is a theoretically and empirically well-motivated framework
for predicting the statistics of the nonlinear matter distribution in
the Universe.  However, current incarnations of the halo model suffer from two major deficiencies:
$(i)$ they do not enforce the stress-energy
conservation of matter; $(ii)$ they are not guaranteed to recover 
exact perturbation theory results on large scales.  Here, we provide a
formulation of the halo model (``\emph{EHM}'') that remedies both drawbacks in a
consistent way, while attempting to maintain the predictivity of the
approach.  In the formulation presented here, mass and momentum conservation are guaranteed on large scales,
and results of perturbation theory and the effective field theory can
in principle
be matched to any desired order on large scales.  We find that a key
ingredient in the halo model power spectrum is the halo stochasticity
covariance, which has been studied to a much lesser extent than other ingredients such
as mass function, bias, and profiles of halos.  As written here, this
approach still does not describe the transition regime between perturbation
theory and halo scales realistically, which is left as an open problem.  
We also show explicitly that, when implemented consistently, halo model
predictions do not depend on any properties of low-mass halos that are smaller than 
the scales of interest.  
\end{abstract}

\date{\today}

\maketitle

\section{Introduction}
\label{sec:intro}

In the halo model (see \cite{cooray/sheth} for a review), all matter in the universe is assumed to be within virialized structures, called halos.  
Under this assumption, the statistics of matter on all scales are determined
by the statistics of these halos as well as their density profiles.  
Most incarnations of the halo model further assume that halos are mutually exclusive, such
that each mass element is part of one and only one halo, and we will do so as well here. 

Currently frequently employed incarnations of the halo model (e.g., 
\cite{
2003MNRAS.344..857T,
2003MNRAS.346..949T,
2004PhRvD..70d3009H,
2008PhRvD..77d3507Z,
2009PhRvD..79h3518S,
2009MNRAS.395.2065T,
2010PhRvD..81f3005S,
2011PhRvD..83d3526S,
2013MNRAS.429..344K,
TakadaHu,
2014JCAP...03..021L,
2014JCAP...04..029B,
li/hu/takada,
2014JCAP...12..053M,
2015MNRAS.452.2247V,
2015PhRvD..92f3517L,
2015arXiv151000398K,
Lazanu/etal:15,
2015MNRAS.454.1958M,
2015MNRAS.454.3030P}) 
have two major deficiencies:  
$(i)$ they do not enforce the physical constraint of stress-energy
conservation of matter; $(ii)$ they are not guaranteed to recover 
exact perturbation theory (PT) results on large scales.  The most widely known symptom
of these deficiencies is the $k$-independent white noise contribution to the matter power 
spectrum $P_{mm}(k)$ of the \emph{1-halo} term on large scales.  
The goal of this paper is to address these issues while attempting to preserve
the successes of the halo model, namely its predictivity: the ability
to provide a reasonably good description of matter and halo statistics over a wide range of scales with few free parameters.  

For this reason, we also demand $(iii)$ \emph{self-consistency}, namely
that the same set of parameters describes 
the nonlinear $n$-point correlations of  matter as well as the
cross-correlations of matter with halos.  
This crucially requires that nonlinear and nonlocal
bias is incorporated in the model.  Further, we demand that the halo model also consistently 
describe the cross-correlation of the nonlinear matter density with the initial conditions; i.e., in case of the power
spectrum, the matter power spectrum $P_{mm}(k)$ and the cross-correlation
(propagator) between the initial density field evolved forward using PT,
and the final, nonlinear density fields, $P_{1m}(k)$.  
The former receives stochastic contributions, while the latter only contains
the \emph{deterministic} terms, i.e. contributions that correlate with initial
perturbations with wavenumbers of order $k$ and smaller.  $P_{mm}(k)$ and $P_{1m}(k)$ can be used to extract the stochastic contribution
to the matter power spectrum in simulations \cite{tassev/zaldarriaga,baldauf/schaan/zaldarriaga}.  

To this end, we describe how the halo model can be constructed consistently
up to a given order, with a finite set of free parameters, so that it 
is guaranteed to satisfy mass and momentum conservation on scales
much larger than individual halos, as well as
matching the exact perturbative solution to the same order on large scales, 
including effective beyond-fluid terms (but see next paragraph).  
In this sense, the halo model consistently extends the predictions of the effective 
field theory (EFT) of LSS \cite{baumann/etal} into the nonlinear regime.    
Of course, on fully nonlinear scales the predictions are not guaranteed
to be correct or within a rigorously calculable theoretical uncertainty of the
correct answer.  Since various implementations of the halo model
paradigm have been presented in the literature, we will adopt
the shorthand \emph{EHM} for the specific construction presented here.  

There is a further well-known trouble with the halo model which we \emph{do not}
address:  in the transition region between scales where perturbation theory
is valid, and scales that are mostly determined by halo density profiles
(1-halo regime), the halo model is known to not describe simulation
results well.  Moreover, the situation becomes worse when
going to higher order in PT.  As the focus of this paper
is on a consistent description of large scales, we will have not much
to say on this here.  It is likely however that the halo model implementation
presented here will need to be extended to solve this issue (see \refsec{Pk1l}).  

Let us briefly describe the relation to previous attempts at resolving the
above mentioned issues of the halo model.  
Note that these attempts were partially motivated by modeling the transition
regime mentioned above.  Refs.~\cite{VN1,VN2} (\emph{Halo-PT}) performed
a separation of matter statistics into $n$-halo terms in Lagrangian space.  
This offers the advantage of a simple implementation of mass conservation.  
On the other hand, one needs to assume a specific exclusion model for halos, and it
is not possible to specify the bias parameters of halos in the model.  Thus,
a consistent connection to  perturbation theory on large scales does not appear to be
feasible in this approach.  
Further, as pointed out in \cite{VN1}, the stochastic contribution to $P_{mm}(k)$
does not scale as $k^4$ in the low-$k$ limit, as required by mass and momentum
conservation, but as $k^2$.  

Refs.~\cite{mohammed/seljak,seljak/vlah} give a prescription (\emph{Halo-Zel'dovich}) for the matter power spectrum
and its covariance based on the power spectrum in the Zel'ldovich approximation, to which a power series in $k^2$ is added.  
The latter can be interpreted as an expansion of the mean halo 
profile.  At low $k$, the lowest order coefficient can be matched to the 1-loop
power spectrum predicted by perturbation theory, in order to achieve the correct
large-scale limit.  For this, \cite{seljak/vlah} had to introduce a compensating
kernel in order to cancel the $k^0$ contribution from the profile expansion.  
Note that the stochastic and deterministic contribution to $P_{mm}(k)$ 
are not modeled separately in this approach, which only considers their sum.  
Since the model is built on the matter power specturm in the Zel'dovich
approximation, nonlinear halo bias is not included in this prescription.

Thus, while these ansatzes recognize the problems of the standard halo model, 
and point to possible approaches to solve the transition regime problem,
they do not satisfy all of the conditions $(i)-(iii)$ mentioned above 
in their current form, since they do not consistently
describe halo correlations and the stochastic contribution to $P_{mm}(k)$.  

The outline of the paper is as follows.  \refsec{HM} describes
the general procedure for constructing EHM, and spells out the
assumptions made; this section constitutes the core of the paper.  
We then describe the lowest order (tree-level) incarnation of EHM and its
prediction for the matter power spectrum in \refsec{Pk}.  The
following \refsec{Pk1l} discusses aspects of the next-to-leading order
(1-loop) EHM power spectrum prediction.  After that,
we consider the tree-level bispectrum in \refsec{Bk}.  
\refsec{vel} contains a brief discussion of the matter velocity field.  
We make some comments regarding the relation to the EFT of LSS
in \refsec{EFT}, before concluding in \refsec{concl}.  The appendices
discuss the issue of the low-mass cutoff, halo triaxiality, and
present the expressions for the bispectrum that are too lengthy for
the main text.  The discussion of the low-mass cutoff in \refapp{lowmass}
is also relevant for other frequently used versions of the halo model.  

\section{General self-consistent halo model}
\label{sec:HM}

In this section we describe the general procedure for relating the
matter density perturbation
\be
\d_m(\vx,\tau) \equiv \frac{\rho(\vx,\tau)}{\rhob(\tau)}-1\,,
\ee
where $\rhob$ is the background density, to halo properties.  We begin
by allowing for fully general halo clustering and profiles.  Afterwards,
we will assemble $\d_m$ and show what constraints mass and momentum
conservation of matter place on the halo properties.

\subsection{Halo clustering}
\label{sec:dh}

Let us begin with the description of the halo density field at fixed
mass $M$.  
In slight abuse of notation, we denote the local number density of halos
per logarithmic mass interval as $n(M,\vx,\tau)$.  This will not
lead to confusion as we will never consider any other type of halo
mass function.  The cosmological
average of the same quantity is defined as $\bar n(M,\tau)$.  The number density
perturbation of halos at a given mass is correspondingly denoted as
\be
\d_{h,M}(\vx,\tau) \equiv \frac{n(M,\vx,\tau)}{\bar n(M,\tau)} - 1\,.
\ee

Let us consider large scales, that is, scales much larger than the Lagrangian
radius $R_L(M)$.  The EHM ansatz we will pursue here assumes that
\emph{all higher derivative terms are supplied by halo profiles.}  
Then, it is sufficient to describe the clustering of halos at lowest
order in derivatives, significantly reducing the number of free parameters
of the model.  Relaxing this assumption is one possibility to address
the failures of EHM in the transition regime (\refsec{Pk1l}) however.  
The equivalence principle guarantees the absence of halo velocity bias 
at lowest order in derivatives \cite{MSZ}.  
In other words, at lowest order in derivatives halos move along the trajectories of the matter fluid itself.  
We can then write, to any given order in perturbation theory,
\ba
\d_{h,M}(\vx,\tau) =\:& \sum_O \left\{ b_O(M,\tau) + \eps_O(M,\vx,\tau)\right\} \left[O\right](\vx,\tau) \vs
& + [\eps](M,\vx,\tau)\,,
\label{eq:bias1}
\ea
where $b_O(M,\tau)$ are bias parameters and $[O]$ are renormalized bias
operators constructed out of the density, tidal field and convective
time derivatives of the same.\footnote{Note that it would be more accurate to write $[\eps_O O]$, since this combination is renormalized jointly.}   Complete bases for the bias expansion have
been described in \cite{MSZ,Angulo:2015eqa}; renormalization of bias
operators is described in \cite{mcdonald:2006,assassi/etal}.  
The fields $\eps,\,\eps_O$ are stochastic
fields with zero means which are completely characterized by their moments
$\<[\eps](M,\vx,\tau)[\eps](M',\vx,\tau)\>$ and so on (again, this holds
at lowest order in derivatives).  
The explicit bias expansion to linear order [\refeq{dhlin}] and
second order [\refeq{dh2nd}] will be given below.  
In general, the bias parameters of halos are not uniquely determined
by their mass, a phenomenon known as assembly bias.   
We will neglect this effect in the main text and discuss it briefly in \refsec{concl}.

\subsection{Halo profiles}
\label{sec:prof}

Further, we will also need a prescription for the density profiles
of halos, which we write as
\be
\rho(\vr,M,\tau ) = M\, y(\vr,M,\tau).  
\ee
We will enforce the following mass constraint for the profile:
\be
\int \rho(\vr,M,\tau)\,d^3 \vr = M \int y(\vr,M,\tau)\, d^3\vr = M\,.
\label{eq:yconstraint}
\ee
In EHM, this constraint is essential in order for \refeq{bias1} to be
consistent, and for exact perturbation theory to be matched on large
scales.  It states that the mass function and bias parameters
defined in \refsec{dh} completely characterize the mass distribution on large scales
(at lowest order in derivatives), while the halo profiles provide the detailed 
distribution on small scales.  We will denote the Fourier transform of
$y(\vr,M,\tau)$ (which is dimensionless) as $y(\vk,M,\tau)$.  

Let us assume a mean spherically averaged profile $y(r,M,\tau)$ (we will generalize this below).  
Besides the mass, the halo profile, averaged over an ensemble of halos
within a finite region, will also depend on the local density and 
tidal field.  In general,
we should perturbatively expand the profiles in the local 
gravitational observables in the same way as the halo abundance [\refeq{bias1}],
where now the bias parameters and stochastic fields become functions of $r$
as well.  This formidable set of free functions can however be reduced by using the fact
that spherically averaged halo profiles are usually well described by a 
single number (apart from the mass);  for example, in case of the NFW profile
\cite{NFW}, the concentration $c$.  Then, it is sufficient to write
$y = y(r,M,\tau,c)$ and expand the concentration in a bias expansion of the type
\refeq{bias1}:
\ba
\frac{c(M,\vx,\tau)}{\bar c(M,\tau)} =\:& 1 + \sum_O \left\{ b^c_O(M,\tau) + \eps^c_O(M,\vx,\tau)\right\} \left[O\right](\vx,\tau) \vs
& \quad + [\eps^c](M,\vx,\tau)\,,
\label{eq:biasc}
\ea
where $\bar c(M,\tau)$ denotes the mean halo concentration.  Assuming that the fractional 
fluctuations in the concentration are much less than one, we can then 
expand
\ba
& y(r,M,\tau,c(\vx,\tau)) = y(r,M,\tau,\bar c) \label{eq:yexp} \\
& + y_c(r,M,\tau,\bar c) 
\left[b_1^c(M,\tau) \d(\vx,\tau) + [\eps^c](M,\vx,\tau) + \cdots \right]
\nonumber
\ea
where
\be
y_c(r,M,\tau, c) \equiv \frac{\partial}{\partial\ln c} y(r,M,\tau, c)\,.
\ee
Note that \refeq{yconstraint} implies that $\int d^3\vr\, y_c(r,M,\tau,c)=0$.  

In general, we should also take into account that halos are triaxial.    
This was investigated in \cite{2006MNRAS.365..214S}.  Moreover,
the orientation of the axes will correlate with large-scale tidal fields.  
We study this in \refapp{triax}, and find that,
under reasonable assumptions, the terms introduced by allowing
for halo triaxiality are degenerate with those obtained through
the isotropic profile expansion \refeq{yexp}.  Thus, we can effectively
account for triaxiality through this expansion.  This is very useful
as it reduces the number of free parameters in the halo model
predictions.  

Let us consider the Fourier transform of the profile on large scales,
i.e. at low $k$.  \refeq{yconstraint} implies that 
$y(k\to 0,M,\tau) = 1$.  Moreover, we can expand
\be
y(k,M,\tau,c) \stackrel{k \to 0}{=} 1 - a_M\, k^2 R_M^2 + \O(k^4)\,,
\label{eq:ylowk}
\ee  
where $R_M$ is the Eulerian halo radius (e.g., $R_{200}$)  and $a_M$
is a mildly mass-dependent number of order one that depends on the exact
profile and mass-concentration relation assumed.  \refeq{ylowk}
will be useful when considering the low-$k$ limit of matter statistics
in the halo model.  Further, we immediately see that
\be
y_c(k,M,\tau,c) \stackrel{k \to 0}{=} - \frac{\partial a_M}{\partial\ln c}\, k^2 R_M^2 + \O(k^4)\,,
\label{eq:yclowk}
\ee  
scaling as $k^2$ in the $k\to 0$ limit.

\subsection{Matter density}
\label{sec:dm}

Following the halo model paradigm, the matter density perturbation $\d_m$ is given by
a superposition of halos weighted by their density profiles.  
Let us denote the frequently appearing mass weighting integral as
\be
\int d\rho(M,\tau) \equiv \int d\ln M\,\frac{M}{\rhob} \bar n(M,\tau)\,.
\label{eq:massint}
\ee
Note that $d\rho$ is dimensionless, and that $\int d\rho(M,\tau) = 1$
in order to satisfy \emph{global} mass conservation at the background level, which corresponds
to the well-known integral constraint on the mass function.\footnote{Note that we do not need to assume that the mass function is universal, i.e. determined by a function $f[\d_c/\sigma(M)]$.}  
\refeq{massint} formally requires a parametrization $\bar n(M,\tau)$ for all 
masses.  We will discuss this issue at the end of this section.  

The fractional matter density perturbation is then in full generality given by
\ba
1+\d_m(\vx) = \int\! d\rho(M) \!\int & d^3 \vy\, 
 \left[1+\d_{h,M}(\vy) \right] \vs
&\times
y[\vx-\vy, M,c(\vy)]\,,
\label{eq:dmHM}
\ea
where here and in the following we drop the explicit time argument for clarity (in the following, we will always work at some fixed time $\tau$).  

\refeq{dmHM} by itself is not sufficient however, since $\d_{h,M}$ in turn is constructed from $\d_m$.  Here, we introduce the following procedure.  
First, one expands $\d_{h,M}$ and $c(\vy)$ to a fixed order in perturbation theory.  
For example, to match PT predictions for the 1-loop power spectrum, we
need to expand $\d_{h,M}$ to third order in perturbations (\refsec{Pk1l}).  For consistency,
one should similarly expand $c$ (around $\bar c$) to third order, unless
those terms are numerically suppressed (see \refsec{Pk}).    
As we noted above, \refeq{yconstraint} ensures that the
terms involving the concentration are higher order in derivatives.  
Then, the desired statistics of $\d_m$ 
are given as convolutions of correlators of the renormalized operators appearing in the bias
expansion \refeq{bias1} [and \refeq{biasc}] with the
halo density profiles.  We will see explicit examples of this in the
following sections.  This approach assumes that all corrections to
the PT matter density field are effectively modeled by the halo profiles $y(\vy,M,c)$.  
We will discuss the issues related to this assumption in \refsec{Pk1l}.

It is important to emphasize again that \emph{all} matter and halo
statistics follow unambiguously from this procedure, so that the same set
of halo properties $\bar n(M),\, b_O(M),\, y(k,M)$ describe all these
observables.  Further, in most studies to date, halo model statistics 
were derived at tree level in perturbation theory.  In EHM,
this is not necessary, and the halo model can be extended to match 
perturbation theory at any desired order.  We will see one example
(and the associated issues) in \refsec{Pk1l}.  

The bias expansion \refeq{bias1} describes the distribution of matter
among halos on large scales, i.e. scales much larger than typical sizes
of halos, independently of their internal structure.  Combining this
with the fact that halos are comoving with matter on large scales,
it is easy to see that \emph{local} mass and momentum conservation
simply imply the following constraints on the bias parameters and
stochasticities:
\ba
\int d\rho(M)\,b_O(M) =\:& \left\{
\begin{array}{ll}
1, & O = \d \\[3pt]
0, & \mbox{otherwise}
\end{array}\right. \label{eq:conscond}\\[3pt]
\int d\rho(M) [\eps](M,\vk) \stackrel{k \to 0}{=}& 0 + \O(k^2) \vs
\int d\rho(M) [\eps_O(M) O] (\vk) \stackrel{k \to 0}{=}&\: 0 + \O(k^2)\,,
\nonumber
\ea
which hold at all times.  
The first line states that the mass-weighted mean linear bias of halos 
should be 1, while the corresponding mean bias vanishes for all nonlinear
terms.  The conditions on $\eps,\,\eps_O$ are to be understood as 
constraints on the auto- and cross-correlations between the renormalized stochastic
fields in the low-$k$ limit.  That is, they imply for example
\ba
\int d\rho(M) \int d\rho(M') \Big\langle [\eps_O(M) O](\vk)& [\eps_{O'}(M') O'](\vk') \Big\rangle' \vs
\stackrel{k\to 0}{=}& \O(k^4)\,.\label{eq:Peps1lgen}
\ea
One can also interpret the stochasticity constraints locally however:  
if we consider the matter density at a given point, coarse-grained 
on a sufficiently large scale (much larger than the radius of typical halos), 
then the stochasticity of halos of various mass cancels after mass weighting.  
That is, there might be more halos at some fixed mass in a given realization of
initial phases, but this has to be compensated by a smaller numer of
halos at other masses such that the total amount of matter is locally
conserved.  

Apart from ensuring mass and momentum conservation, these conditions are sufficient to ensure that on scales larger than halos,
the matter density \refeq{dmHM} reduces to the perturbation theory
prediction at the desired order.  
The constraint on the stochasticity will become particularly relevant
in the following sections, as it is responsible for removing the constant
tail of the standard 1-halo term in the low-$k$ limit.  
Naturally, any constraints that can be placed on the opposite, small-scale
limit are very useful as anchor points.  First, most obviously, one can use the existing, very
accurate measurements of mean halo profiles.  Second, we can also place
physical constraints on the stochastic terms in the high-$k$ limit.  

Before turning to this limit, let us discuss another issue
related to \refeq{conscond}.  The integral $\int d\rho(M)$
formally extends to arbitrarily small halo masses, far beyond the range that is empirically calibrated
with simulations.  In fact, for standard parametrizations of the mass function,
the mass-weighting integrals in \refeq{conscond} typically
converge very slowly towards low masses.  
Since properties of very low-mass halos are poorly constrained by simulations, this raises
the question of whether the halo model predictions discussed here
and presented in the literature actually rely
on extremely low-mass halos whose properties are poorly known?  

Fortunately, as we show in \refapp{lowmass}, the answer is no.  Specifically,
if the properties of halos are calibrated to a minimum mass $M_s$, then
one can cut off the mass-weighting integral below $M_s$, and introduce
compensating parameters to enforce the conditions in \refeq{conscond}.  
After this procedure,
any systematic uncertainties  in the halo model predictions due to the
mass cut scale as $(k R_{M_s})^2$.  These systematics reach 10\% at a scale of
\be
k_{10\%} \approx 5.6 \iMpch \left(\frac{M_s}{10^{10} \Msunh} \right)^{-1/3}\,.
\ee
Given the current advanced state of high-resolution simulations, this
is not likely to be an important constraint for cosmological applications
of the halo model.  Note that the procedure we describe in \refapp{lowmass}
applies to any halo model prescription that involves integrals over
halo masses.

\subsection{Stochasticity in the high-\texorpdfstring{$\bm{k}$}{k} limit}
\label{sec:stochhighk}

The stochastic terms $\eps,\, \eps_O$ are non-perturbative
and numerically important in the high-$k$ limit.  
For scales much smaller
than the sizes of halos (of a given mass), the stochasticity in the halo
abundance should approach Poisson statistics governed by the \emph{local}
halo abundance $\bar n(M)[1 + \d_{h,M}]$.  This is because Poisson
statistics apply if the
wavelength $1/k$ of a given mode is much smaller than the mean 
inter-halo separation, that is, if $\bar n/k^3 \ll 1$.  Further,
since halos are non-overlapping in the halo model (each matter particle
only belongs to one parent halo), halos of different mass have independent
Poisson noise.  These are significant
constraints, since in this limit, it completely determines the moments of
$[\eps]$ as well as all $[\eps_O]$.  This works as follows.  For clarity,
we will drop the brackets around $\eps,\,\eps_O$ in the remainder of this 
section, keeping in mind that we always deal with the renormalized fields.

Consider halos within an infinitesimal logarithmic mass interval $d\ln M$ centered around a fixed mass $M$, and a fictitious small volume element $V$ around point $\vx$
such that 
\be
\bar N \equiv V \bar n\, d\ln M \ll 1\,.
\ee
The Poisson assumption states that
the halo number within this volume follows a Poisson distribution,
\ba
N(\vx) \sim {\rm Poisson}\left[\bar N \left(1 + \sum_O b_O [O](\vx) \right) \right] \,.
\label{eq:stochpoisson}
\ea
Here, the operators $[O](\vx)$ are considered to be coarse-grained on some larger
scale (of order the halo radius, for example).  
We can subtract the mean, which corresponds to the deterministic part
of the bias expansion \refeq{bias1}, and call the remainder $\bar N\,\eps_p(\vx)$ 
with $\<\eps_p\> = 0$. \refeq{stochpoisson} then specifies the moments
of $\eps_p$, i.e.
\ba
\< \eps_p^2 \> =\:& \frac1{\bar N} \left(1 + \sum_O b_O [O](\vx) \right) \vs
\< \eps_p^3 \> =\:& \frac1{\bar N^2} \left(1 + \sum_O b_O [O](\vx) \right)\,,
\label{eq:Pmoments}
\ea
and so on.  On the other hand, we have a specific perturbative
expansion of the stochasticity in \refeq{bias1}, which yields
\be
\eps_p(\vx) = \eps(\vx) + \sum_O \left[\eps_O O\right](\vx)\,.
\label{eq:Pmoments2}
\ee
By matching \refeq{Pmoments2} to the moments derived from \refeq{Pmoments}, 
and using the fact that there is only a single random field $\eps_p$ (at fixed halo mass), we can then
uniquely determine the moments of $\eps$ and $\eps_O$, order by order.  
Performing a Fourier transform within the volume $V$, we then obtain
the desired high-$k$ limit of the moments in Fourier space.  
For example, at linear order we simply have
\be
\<\eps(M,\vk) \eps(M',\vk') \>' \stackrel{k\to\infty}{=} \frac{\d_D(\ln M - \ln M')}{\bar n(M)}\,,
\label{eq:epsP2}
\ee
where a prime denotes that the momentum conserving delta function has 
been dropped.  At second order, we obtain the following two additional
constraints:
\ba
&\<\eps(M,\vk) \eps(M',\vk') \eps(M'',\vk'') \>' \stackrel{k\to\infty}{=} 
\label{eq:epsP3}\\
& \hspace*{2.7cm} 
\frac{\d_D(\ln M - \ln M')\d_D(\ln M - \ln M'')}{[\bar n(M)]^2} \vs
& \< \eps(M,\vk) \eps_\d(M',\vk') \>' \stackrel{k\to\infty}{=} 
\frac12 b_1(M) \frac{\d_D(\ln M - \ln M')}{\bar n(M)}\,.
\nonumber
\ea
These are all stochastic moments that exist at second order (\refsec{Bk}).  
The second line of \refeq{epsP3} is directly related to the
\emph{halo sample variance} disussed in \cite{2013MNRAS.429..344K}.  
Note that both \refeq{epsP2} and \refeq{epsP3} violate the
constraints \refeq{conscond} in the opposite, large-scale limit.  
This already shows that a scale-dependent stochasticity is a 
necessary part of a consistent formulation of the halo model.  
The entire reasoning of this section also applies to the stochastic
fields appearing in the profile expansion \refeq{biasc}.  Moreover,
in the high-$k$ limit these fields are uncorrelated with the
stochasticity in the halo number.

\section{Lowest order halo model and power spectrum}
\label{sec:Pk}

The lowest-order consistent incarnation of the halo model expands \refeq{bias1} to linear order, 
\ba
\d_{h,M}(\vx) =\:& b_1(M) \d_1(\vx) + [\eps](M,\vx)\,,
\label{eq:dhlin}
\ea
where $\d_1$ denotes the linear density field.  
In addition, the profiles are expanded via \refeq{yexp}:
\ba
\label{eq:clin}
 y(r,M,c(\vx)) =\:& y(r,M,\bar c) \\
& + \left[b_1^c(M) \d_1(\vx) + [\eps^c](M,\vx) 
\right] y_c(r,M,\bar c) \nonumber
\ea
The matter density perturbation is then given as a mass-weighted integral
of the halo number density convolved with the halo density profile as
in \refeq{dmHM}.  

Let us first look at the matter propagator, i.e. the cross-correlation of $\d_m$ with the PT-evolved density field (here just the linear density field) in Fourier space:
\ba
P_{1m}(k) =\:& \int d\rho(M)
\left[b_1(M) y(k,M) + b_1^c(M)  y_c(k,M) \right] \vs
& \qquad\times P_{\rm L}(k)\,,
\label{eq:Pkhm}
\ea
where here and in the following we will drop the explicit concentration
argument when it is set to the mean value $\bar c(M)$, and $P_{\rm L}$ denotes
the linear matter power spectrum.  Using the
low-$k$ behavior of $y$ and $y_c$ and \refeq{conscond}, we see that
in the low-$k$ limit we recover
\be
P_{1m}(k) = P_{\rm L}(k) \left[ 1 + \O( R_{\rm HM}^2 k^2 ) \right]\,,
\label{eq:Pm1lowk}
\ee
where 
\be
R_{\rm HM}^2 \equiv \int d\rho(M)\, a_M b_1(M) R_M^2\,.
\ee
This is the characteristic scale that appears in the low-$k$ limit
of EHM, and is of order $R_{M_*}$, where $M_*$ is defined through 
$\sigma(M_*) = \d_c$;  that is, $R_{\rm HM}$ is of order the typical Eulerian halo radius.  
Note that this scale is smaller than the nonlinear scale $1/k_{\rm NL}$ where the
density contrast becomes of order 1.

We now turn to the matter power spectrum.  This is given by
\ba
& P_{mm}(k) = \int d\rho(M)\int d\rho(M') \  \Bigg\{  \label{eq:Pkmm}\\
& y(k,M) y(k,M') \Big[
b_1(M) b_1(M') P_{\rm L}(k) 
+ P^{\eps\eps}_{MM'}(k)
\Big] \vs
+\:&   2 y(k,M) y_c(k,M') \Big[ b_1(M) b_1^c(M') P_{\rm L}(k) 
+ P^{\eps\eps^c}_{MM'}(k)
\Big]
\vs
+\: &  y_c(k,M) y_c(k,M') 
\Big[
b^c_1(M) b^c_1(M') P_{\rm L}(k) + P^{\eps^c \eps^c}_{MM'}(k) 
\Big] \Bigg\}
 \,,\nonumber
\ea
where we have defined
\be
P^{\eps^a\eps^b}_{MM'}(k) \equiv \< [\eps^a](M,\vk) [\eps^b](M',\vk') \>'\,.
\label{eq:Pepsgendef}
\ee
\refeq{Pkhm} and \refeq{Pkmm} differ from the standard halo model
power spectrum in two respects: the stochasticity covariances
$P^{\eps^a\eps^b}_{MM'}(k)$;  and the terms from the expansion of halo
concentration in long-wavelength perturbations, proportional to $y_c$.  We will examine both of them in the following sections.  

First however, we consider the low-$k$ limit of \refeq{Pkmm}.  The constraints \refeq{conscond} imply that
\ba
\int d\rho(M) P^{\eps}_{MM'}(k \to 0) =\:& 0 + \O(k^2) \vs
\int d\rho(M) \int d\rho(M') P^{\eps}_{MM'}(k \to 0) =\:& 0 + \O(k^4)\,.
\label{eq:stochcond}
\ea
For the cross-correlation between $\eps,\eps^c$ on the other hand, we only demand $\int d\rho(M) \int d\rho(M') P^{\eps\eps^c}_{MM'} = \O(k^2)$.    
The first line here says that in the low-$k$ limit, the halo stochasticity covariance has
(at least) one zero eigenvalue, with the corresponding eigenvector
given by mass weighting (see also \cite{2009PhRvL.103i1303S,hamaus/etal:2010,2012PhRvD..86j3513H}).  
Ref.~\cite{hamaus/etal:2010} performed a detailed 
analysis in simulations.  Indeed, they find that  
the lowest eigenvalue of $P^{\eps\eps}_{MM'}$ is significantly lower than the 
shot noise $1/\bar n(M)$ of halos in the mass range they considered.  
Further, the corresponding eigenvector is close to mass weighting.  
Similar results were found in \cite{cai/etal:11}.  

Using that $y(k,M) \to 1$ for $k \to 0$, we then see that $P_{mm}(k)$ 
has the same low-$k$ behavior \refeq{Pm1lowk} as $P_{1m}(k)$.  
Moreover, the stochastic contributions, i.e. all terms that involve
$P^{\eps\tilde\eps}_{MM'}$, scale as $k^4$ in the low-$k$ limit, just
as demanded by mass and momentum conservation.  
The leading contribution to $P_{mm}(k)$ is then
\ba
P_{mm}(k) =\:& P_{\rm L}(k) 
+ \O(R_{\rm HM}^2 k^2) P_{\rm L}(k) \vs
& + \O[k^4,\  k^4 P_{\rm L}(k)]\,.
\label{eq:Pdetk2}
\ea
Note that the 1-loop matter power spectrum contributes terms that also
scale as $k^2 P_{\rm L}(k)$, but involve $1/k_{\rm NL}$ instead of $R_{\rm HM}$.  
This shows that one needs to carry out 
the ``halo model at 1-loop'', by extending \refeq{dhlin} to third order,
in order to obtain a consistent matching to beyond-perfect-fluid terms
in the EFT.

\reffig{Pk} (red solid line) shows the deterministic contribution from the
expansion of the halo density, i.e. the first term in the second line of
\refeq{Pkmm}.  This is the standard \emph{2-halo} term.  Given the discussion
in the previous paragraph, we do not expect this to be a good match
to the true nonlinear power spectrum from simulations.  
For our numerical results, we assume a flat
$\Lambda$CDM cosmology with cosmological parameters given by $\Omega_m=0.27$, $h=0.7$, $\Omega_b h^2=0.023$, $n_s=0.95$, $\sigma_8=0.791$.  
We use the Sheth-Tormen mass function \cite{Sheth/Tormen:1999} and
associated linear bias, and the concentration-mass relation of
\cite{Buletal01}.  We assume that halo masses are given in terms of
a mean interior density equal to the virial density $\rho_{\rm vir} = 363\,\rhob$ for this cosmology.  All results will be shown for $z=0$.  
 
\subsection{Halo stochasticity}

The second term in the second line of \refeq{Pkmm} is the halo model
prediction for the stochastic part of the matter power spectrum.  Here,
with stochastic we mean that it does not correlate with the initial
conditions on the scale $k$ (or at larger scales).  In the halo model,
this is controlled by the halo stochasticity covariance $P^{\eps\eps}_{MM'}(k)$.  
Clearly, this is a key ingredient
of the halo model, as important though much less well studied than the mass function, linear bias, and halo profiles.  The standard halo model assumes
a $k$-independent diagonal covariance following Poisson statistics,
\be
P^{\eps\eps,\,\rm std}_{MM'} = \frac{\d_D(\ln M-\ln M')}{\bar n(M)} \,,
\label{eq:stochhighk}
\ee
which, following \refsec{stochhighk}, is only justified in the high-$k$ limit,
i.e. well within halos.  Moreover, it clearly does not satisfy \refeq{stochcond} and thus
violates mass and momentum conservation.  
Thus, we need to come up with a more physical parametrization of $P^{\eps\eps}_{MM'}$ at low-$k$, which asymptotes to \refeq{stochhighk} if $k$ is larger than the mean inter-halo separation, which is directly related to the Lagrangian radii $R_L(M)$, $R_L(M')$, respectively.

\begin{figure}[t!]
\centering
\includegraphics[width=0.49\textwidth]{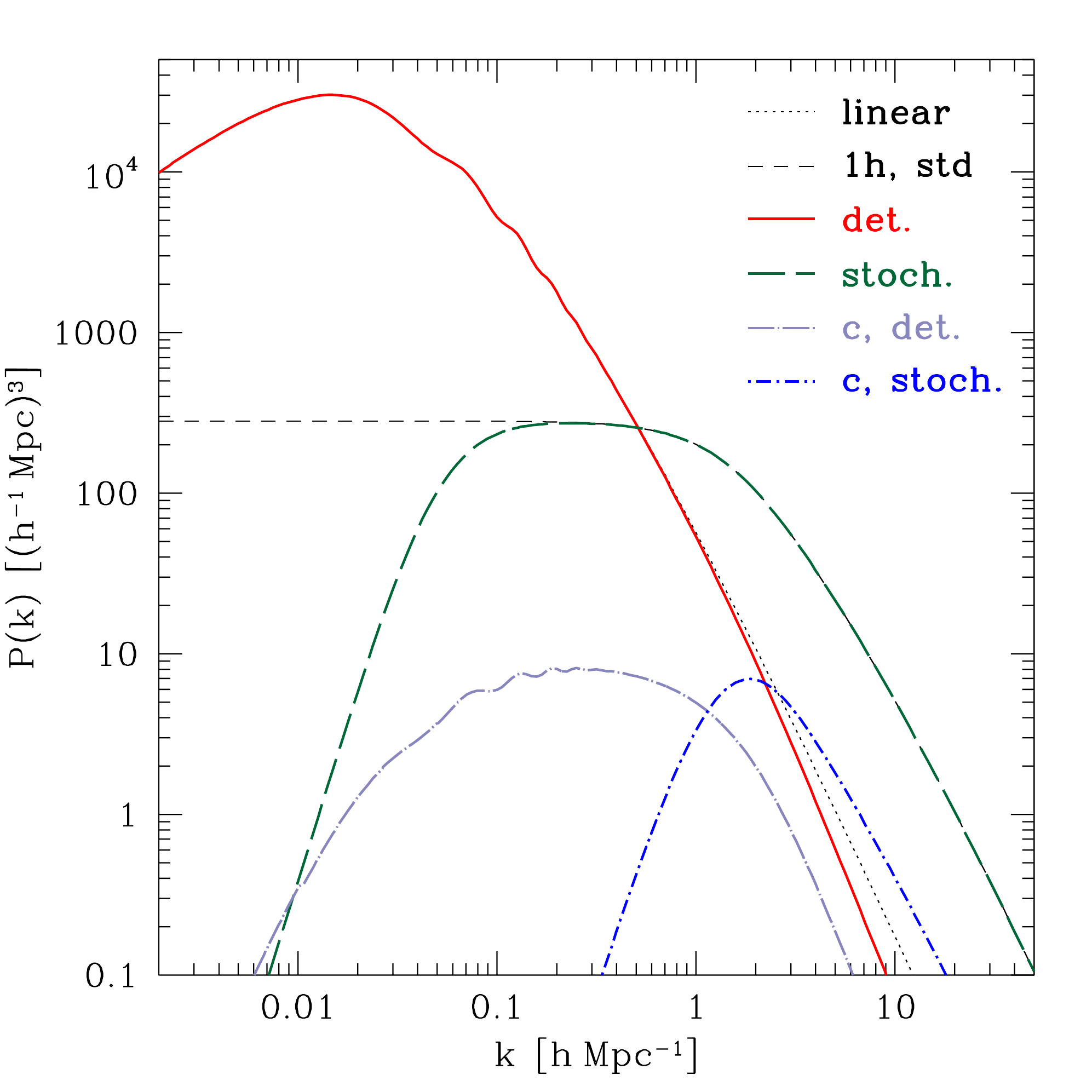}
\caption{Contributions to the lowest-order halo model matter power spectrum 
\refeq{Pkmm} at $z=0$: the red solid line shows the deterministic contribution
[first term in the second line of \refeq{Pkmm}], i.e. the standard 2-halo term,  
while the green long-dashed line is the stochastic contribution
(second term in the same line).  For comparison, we also show the standard
1-halo term as black thin short-dashed line.  The light blue dot-long-dashed
line shows the deterministic contributions from the concentration expansion
[first terms in the third and fourth lines of \refeq{Pkmm}].  Finally,
the stochastic terms of the concentration expansion [second terms in
the same lines of \refeq{Pkmm}] are shown as blue dot-dashed.  
The linear power spectrum is shown as thin dotted line.  
\label{fig:Pk}}
\end{figure}

One possibility is to simply subtract the trace to ensure a zero eigenvalue corresponding
to mass weighting: 
\be
P^{\eps}_{MM'}(k) = \frac{\d_D(\ln M-\ln M')}{\bar n(M)} - \Theta_{MM'}(k) \frac{M\,M'}{\rhob \Mbar} \,,
\label{eq:Peps1}
\ee
where $\Mbar$ is defined as
\be
\Mbar \equiv \int d\rho(M)\,M\,.
\label{eq:Mbar}
\ee
$\Theta_{MM'}(k)$ is an interpolating function that asymptotes to $1$ for $k\to0$,
satisying \refeq{stochcond}, while approaching zero in the high-$k$ limit.  
To be specific, we will choose
\be
\Theta_{MM'}(k) = [1 + (k [R_L(M)+R_L(M')]/2)^4 ]^{-1}\,,
\label{eq:finterp}
\ee
where the transition scale is given by the halo Lagrangian radii following our considerations above.  Note that, for a covariance of the form \refeq{Peps1},
we need $\Theta_{MM'}(k)$ to scale as $1 + \O(k^4)$ in the low-$k$ limit in order
to satisfy the conditions \refeq{conscond} for all $k$.   While different forms of interpolating function could be chosen, we expect the transition to be related to $R_L$.  The detailed shape of the interpolation is not expected to have a significant impact on the power spectrum prediction, as the transition happens on scales where the power spectrum is still dominated by the deterministic contribution (see \reffig{Pk}).  

\refeq{Peps1} is certainly not the only possible choice.  For example,
\cite{hamaus/etal:2010} derived a covariance given by\footnote{Note that this was derived using a standard halo model ansatz based on \refeq{stochhighk}
which does not enforce mass and momentum conservation.  
}
\ba
P^{\eps}_{MM'}(k\to 0) =\:& \frac{\d_D(\ln M-\ln M')}{\bar n(M)} - b_1(M) \frac{M'}{\rhob}
\label{eq:PepsHM}\\
& - b_1(M') \frac{M}{\rhob}  + b_1(M) b_1(M') \frac{\Mbar}{\rhob}\,.
\nonumber
\ea
It is easily verified that this ansatz indeed satisfies 
\refeq{stochcond}, assuming the first line in \refeq{conscond} holds.  
We could thus multiply the last three terms by the interpolating function
$\Theta_{MM'}(k)$ and insert into \refeq{Pkmm}.  
However, the additional terms [in particular the last term in \refeq{PepsHM}]
grow rapidly towards high $k$ due to their mass-weighting, so that
they dominate the matter power spectrum for $k\gtrsim 0.5 \iMpch$ despite the suppression by the interpolating
function \refeq{finterp}; this result is insensitive to the shape 
and steepness of $\Theta_{MM'}(k)$.  Thus, we cannot attain our desired high-$k$
limit, which is the standard 1-halo term based on \refeq{stochhighk}.  We will instead work with \refeq{Peps1} here,  
but conclude that simulation measurements of halo stochasticity on large and 
intermediate scales are essential in order to properly calibrate the halo model prediction.  

The stochastic contribution to the matter power spectrum [second term in the second line
of \refeq{Pkmm}] is shown as green long-dashed line in \reffig{Pk}.  We also show the standard 1-halo prediction with its
unphysical $k^0$ behavior at low $k$ (black short-dashed).  As expected, \refeq{Peps1}
yields the desired $k^4$ behavior of the stochastic contribution, while
asymptoting to the standard 1-halo term for $k \gtrsim 0.5 \iMpch$.  Qualitatively, this is 
what one expects for the stochastic contribution in the halo model, although the
quantitative behavior for $k \lesssim 1\iMpch$ can of course be modified significantly
by changing the low-$k$ limit of $P^{\eps\eps}_{MM'}$ and/or the interpolating function
\refeq{finterp}.  The result appears roughly consistent with the findings
of \cite{baldauf/schaan/zaldarriaga} (e.g., blue curve in Fig. 8), who isolated
the stochastic contribution to the power spectrum in simulations by subtracting the
part correlated with long-wavelength correlations.  Interestingly, they find a slightly shallower
scaling with $k$ than $k^4$ even for $k \lesssim 0.1 \iMpch$.  Whether this really
implies the existence of another scale much below $k_{\rm NL}$ remains to be seen.  

At this point, it is also worth discussing the usual 1-halo vs 2-halo
separation.  The first term in the second line of \refeq{Pkmm} corresponds
to the standard 2-halo term.  One could refer to the second, stochastic term,
as 1-halo term, even though it involves a covariance between different
halo masses.  Alternatively, one could only refer to that part of
the stochastic contribution that is proportional to $\d_D(\ln M-\ln M')$ as
1-halo contribution, while the remainder of the stochastic part is considered
a contribution to a modified 2-halo term (see also \cite{2011PhRvD..83d3526S,baldauf/etal:13}).  
In any case, this separation is
somewhat arbitrary and a matter of definition, as everything should be 
derived from the physical assumptions described in \refsec{HM} rather than
a separation of the statistics into $n$-halo terms.

\subsection{Concentration expansion}

Let us now turn to the terms in the third and fourth line of \refeq{Pkmm},
involving $y_c$, which come from the perturbative expansion of the 
concentration $c$.  First, consider the
deterministic terms $\propto P_{\rm L}(k)$.  
\reffig{Pk} shows these terms, assuming $b_1^c = b_1$ which is almost
certainly a significant overestimation of the effect, given the fairly small
environment dependence observed for the halo concentration in simulations
\cite{Maccio/etal:07}.  In fact, this contribution is entirely dominated
by the cross-term given on the third line of \refeq{Pkmm}.  
Clearly, this contribution is significantly
smaller and shifted to higher $k$ compared to the terms from the
expansion of $\d_{h,M}$.  The main reason for this is that the
change of halo profiles due to a change in concentration happens on fairly 
small scales, of order the scale radius of these halos.  Further, $|y_c(k,M)|$ 
is at most $\sim 0.4$, that is halo profiles do not respond strongly
to a change in concentration.  

Turning to the stochastic terms, 
we now need a parametrization of $P^{\eps\eps^c}_{MM'}(k)$, 
scaling as $\O(k^2)$ for $k\to 0$, and $P^{\eps^c\eps^c}_{MM'}$, which has 
no low-$k$ constraint.  Let us consider the latter.  The simplest
assumption to make is that each halo's concentration is drawn from
a log-normal distribution with fixed scatter $\s_{\ln c}$ around the
mean relation $\bar c(M)$.  Then, we have
\be
P^{\eps^c\eps^c}_{MM'} = \frac{\s_{\ln c}^2}{\bar n} \d_D(\ln M-\ln M')\,.
\ee  
The result, using $\s_{\ln c}=0.4$ (of the order of what was found
for the scatter in concentration in \cite{Maccio/etal:07}), is
also shown in \reffig{Pk}.  Again, we find this to be a small contribution
to $P_{mm}(k)$, mainly relevant around $k \sim 2\iMpch$.  
The final remaining term is the stochastic cross-correlation between halo number
density and concentration $P^{\eps\eps^c}_{MM'}(k)$.  This is expected to
be smaller than the stochastic auto-correlations of halo number and profiles,
because it is constrained to vanish on both small and large scales:  
mass conservation implies a $k^4$ scaling for $k\to 0$, while for
scales $k \gtrsim 1/R_M$ within halos, profiles and number density have
to be independent random variables.  This means $P^{\eps\eps^c}_{MM'}(k)$ can
only be relevant on a fairly narrow range of scales around 
$k\sim 1/(R_M+R_M')$.  For this reason, we do not investigate this term
further here.  

In summary, in the case of the simple concentration expansion of halo profiles performed
here, the effects are suppressed compared to the expansion of
$\d_{h,M}$, so that, depending on the application and range of wavenumbers
of interest, they can be neglected.  We stress however
that this assumes that the impact of the large-scale environment on
halo profiles is well captured by a change in concentration.  If in
reality there is a significant effect on the outer regions of halo profiles,
then this could make the power spectrum contributions from the profile expansion 
more significant and push them to larger scales.  This is well worth
investigating in simulations.  We leave this to future work.

\section{Matter power spectrum beyond tree level}
\label{sec:Pk1l}

The previous section described the leading order prediction of the
halo model, which only matches linear perturbation theory on large scales.  
Let us now turn to the next higher order incarnation of EHM, where we go
to third order in PT.  Our goal is to outline the
overall features of the result, and highlight open issues.  We will
neglect the terms arising from the expansion of the concentration throughout
this section.

Let us begin with the deterministic contributions to the matter power
spectrum.  These can be written as
\ba
& P_{mm}(k)\Big|_{\rm det.} = \int d\rho(M)\int d\rho(M') \, y(k,M) y(k,M') \, \bigg\{  \vs
&  \times b_1(M) b_1(M') \left[ P_{\rm L}(k) + P_{\olp}(k) \right] 
%
+ P_{\rm nlb}^{MM'}(k) \bigg\} \,,
\label{eq:Pkmm1l}
\ea
where 
\be
P_{\olp}(k) = \< \d^{(2)}(\vk)\d^{(2)}(\vk)\>'
+ 2 \< \d^{(1)}(\vk)\d^{(3)}(\vk)\>'
\label{eq:P1l}
\ee
is the one-loop matter power spectrum \cite{Bernardeau/etal:2002}, 
and $\d^{(n)}$ denotes the matter density at $n$-th order in PT.  
$P_{\rm nlb}^{MM'}(k)$ contains
all nonlinear bias terms that are relevant at one-loop order:
\be
P_{\rm nlb}^{MM'}(k) = \sum_{\{O,O'\}\neq\{\d,\d\}}^{\olp} \!\!\!
b_O(M) b_{O'}(M') \< [O](\vk) [O'](\vk') \>'\,.
\label{eq:Pnlb}
\ee
The full expression for the 1-loop halo power spectrum can be found
in \cite{mcdonald/roy:2009,saito/etal:14}.  In analogy with \refeq{P1l},
these terms can be divided into quadratic bias terms which scale similarly
to $\< \d^{(2)} \d^{(2)}\>$, and cubic bias terms which scale similarly
to $\< \d^{(1)} \d^{(3)}\>$.  The numerically largest term of the former
category is given by
\ba
& b_1(M) b_2(M') \< \d^{(2)}(\vk) [\d^2](\vk') \>' \label{eq:Pb2d}\\
& = b_1(M) b_2(M')\int \frac{d^3 \vq}{(2\pi)^3} F_2(\vq,\vk-\vq) P_{\rm L}(\vq) P_{\rm L}(\vk-\vq)\,,\nonumber
\ea
where $F_2$ is the symmetrized second order perturbation theory kernel
\cite{Bernardeau/etal:2002}.  We will use the second-order bias
derived from the Sheth-Tormen mass function for our results [note that this satisfies
\refeq{conscond}].  
\begin{figure}[t!]
\centering
\includegraphics[width=0.49\textwidth]{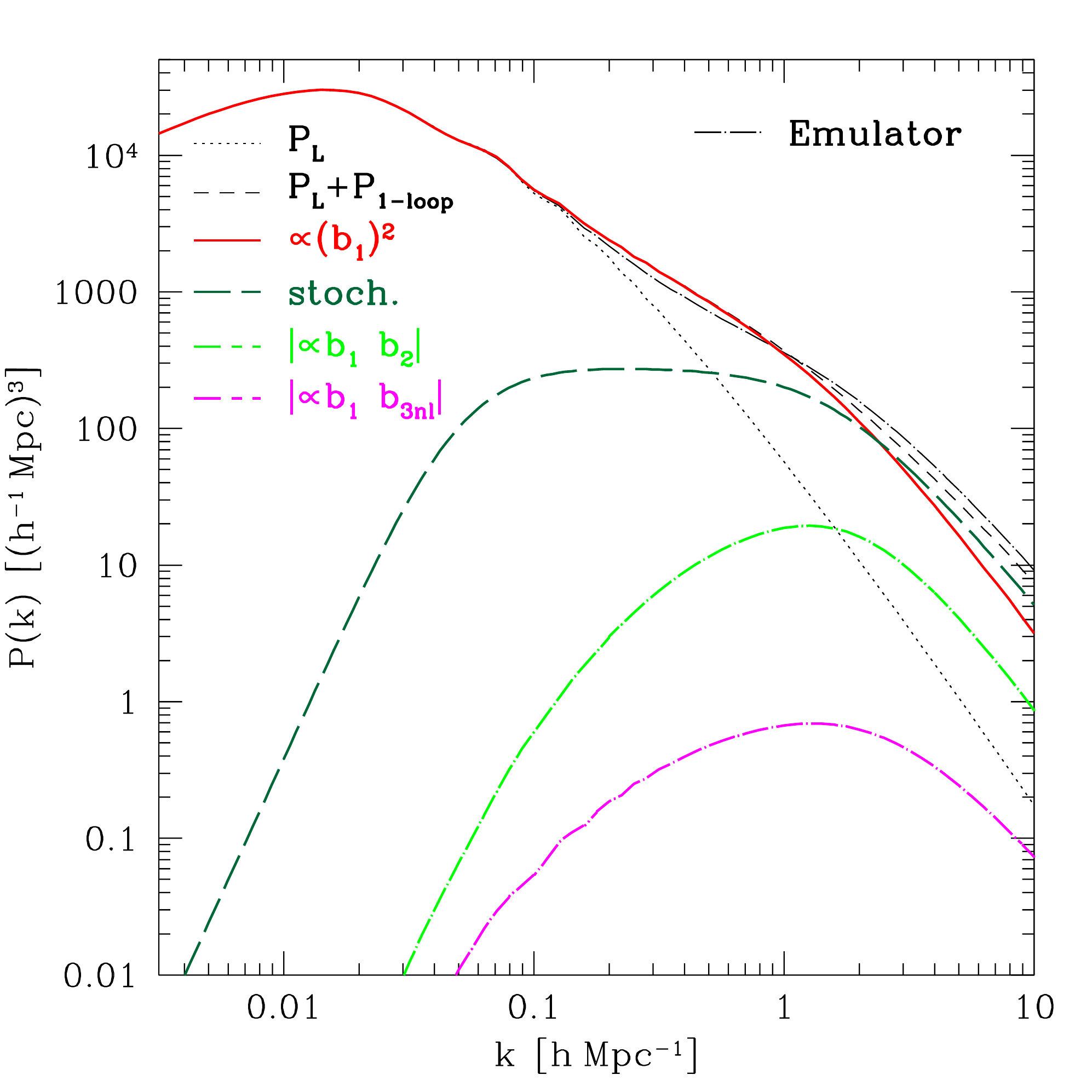}
\caption{Illustration of some of the contributions to the 1-loop halo model matter power spectrum \refeq{Pkmm1l} at $z=0$: the red solid line shows the contribution
$\propto b_1(M) b_1(M')$ (terms in brackets in the second line of \refeq{Pkmm1l}).  The green long-dashed line is the same stochastic contribution
as in \reffig{Pk}.  The light blue dot-long-dashed
line shows the contribution from \refeq{Pb2d}, while the third order bias
contribution [\refeq{Pb3nl}] is shown as blue dot-dashed.  
The linear power spectrum is shown as thin dotted line as in \reffig{Pk},
while the thin dashed line shows the matter power spectrum at 1-loop.  
The thin dot-long-dashed shows the nonlinear power spectrum from the
Coyote emulator \cite{emulator}.  
\label{fig:Pk1l}}
\end{figure}
At 1-loop order, the third order
renormalized bias contributions to \refeq{Pnlb} are all degenerate and 
can be grouped as a single contribution \cite{mcdonald/roy:2009}:
\ba
 \sum_{O\,=\,\O(\d^3)} & b_1(M) b_O(M') \< \d^{(1)}(\vk) [O](\vk) \>' \vs
& = b_1(M) b_{\rm 3nl}(M') \sigma_3^2(k) P_{\rm L}(k) \,,
\label{eq:Pb3nl}
\ea
where $\sigma_3^2(k)$ is a filtered version of the linear power spectrum.  
The filter is defined explicitly in \cite{mcdonald/roy:2009}.  
For illustratitive results, we use the prediction from local Lagrangian
biasing \cite{saito/etal:14},
\be
b_{\rm 3nl}(M) = \frac{32}{315} [ b_1(M)-1 ]\,.
\ee
Let us also give the expression for the 1-loop propagator, i.e. the cross-correlation of the
PT-evolved initial density field and the nonlinear matter density:
\ba
 P_{1m}(k) = & \int d\rho(M) \, y(k,M)  \, \bigg\{ 
b_1(M) \left[ P_{\rm L}(k) + P_{\olp}(k) \right] \vs
& + \sum_{O\,=\,\O(\d^2)} b_O(M) \< \d^{(2)}(\vk)[O](\vk)\>' \vs
& + \sum_{O\,=\,\O(\d^3)} b_O(M) \< \d^{(1)}(\vk)[O](\vk)\>' \bigg\}\,,
\label{eq:Pk1m1l}
\ea
where the leading contributions to the second and third line are given by
\refeqs{Pb2d}{Pb3nl} without the factor $b_1(M)$.    

Let us finally turn to the stochastic terms at 1-loop, given by
\ba
 P_{mm}&(k)\Big|_{\rm stoch.} = \int d\rho(M)\int d\rho(M') \, y(k,M) y(k,M') \vs
  \times& \bigg[ P^{\eps\eps}_{MM'}(k)
+ \int \frac{d^3 \vq}{(2\pi)^3} P^{\eps_\d \eps_\d}_{MM'}(q) P_{\rm L}(|\vk-\vq|) \vs
& \  + C_1 + C_2\,k^2  \bigg]\,.
\label{eq:Pkst1l}
\ea
There is only loop additional contribution to the matter power spectrum
which involves $P^{\eps_\d \eps_\d}$ defined following
\refeq{Pepsgendef}. In order to enforce \refeq{Peps1lgen},
we add counterterms $C_1$ and $C_2$, whose values are uniquely 
determined given $P^{\eps_\d \eps_\d}_{MM'}(k)$.  These counterterms
ensure that the final contribution scales as $k^4$ in the low-$k$ limit.  
While evaluating this term requires a parametrization of $P^{\eps_\d \eps_\d}_{MM'}$,
we have performed a rough evaluation using a form inspired by the
high-$k$ limit discussed in \refsec{stochhighk}.  
Including the counterterms, this contribution was found to be of order a few
percent of the tree-level stochastic term for $k\lesssim 0.5\iMpch$.  
Given the lack of knowledge
about $P^{\eps_\d \eps_\d}_{MM'}$ on large and intermediate scales, we do not
show it here.  

\reffig{Pk1l} shows the contribution $\propto b_1(M) b_1(M')$ in 
\refeq{Pkmm1l}, as well as the two terms \refeq{Pb2d} and \refeq{Pb3nl}
which are a subset of $P_{\rm nlb}^{MM'}(k)$.  Note that, after mass weigthing,
both \refeq{Pb2d} and \refeq{Pb3nl} yield negative contributions to \refeq{Pkmm1l}.  
We also show the linear and 1-loop matter power spectra as well as the tree-level
stochastic term.  It is clear that the latter together with the
term scaling as $b_1(M) b_1(M')$ dominate the EHM
power spectrum.  The nonlinear biases are suppressed by the
conservation conditions \refeq{conscond}, so that they only begin
to contribute on scales of order the halo radius (where $y(k,M)$ 
begins to be appreciably different from 1).  This suppression
is even stronger for terms that scale as $(b_2)^2$.  

Thus, although \reffig{Pk1l} does not show all EHM contributions, we can already
draw some conclusions.  For comparison, we also show in \reffig{Pk1l} the nonlinear matter
power spectrum evaluated for our fiducial cosmology by the 
Coyote emulator \cite{emulator}, which is accurately calibrated on
simulations.  This illustrates the well-known fact that $P_{\olp}$ overpredicts
the true nonlinear power spectrum measured in N-body simulations.  
The stochastic contribution, at least assuming our parametrization \refeq{Peps1},
only exacerbates this problem.  The fact that some of the nonlinear
bias terms are negative will not solve this issue in the range $k\sim 0.2-1 \iMpch$, 
as they are too small numerically.  

This problem occurs on intermediate scales, which are too small for perturbation theory to be valid, but still larger than the Eulerian radius of
halos.  
Thus, one cannot expect rigorous physical solutions by extrapolating 
from either regime.  Note that the predictions in the intermediate regime
also depend on which perturbative scheme is used, e.g. Eulerian SPT vs
Lagrangian LPT: different schemes are only guaranteed to give the same
result on scales where perturbation theory is valid, and will diverge
on fully nonlinear scales.    

In the framework of the halo model, these issues arise because 
we assume perturbation theory to describe $\d_{h,M}$ correctly down to
halo scales, which does not hold.  
For example, nonperturbative effects such as halo exclusion are not
included by definition.  One approach to address the mismatch on intermediate
scales is to perform a matching to simulations.  In particular, one can match the halo propagator,
which for 1-loop SPT is given in analogy to \refeq{Pk1m1l} by
\ba
 P_{1h}(k,M) =\:& b_1(M) \left[ P_{\rm L}(k) + P_{\olp}(k) \right] \vs
& + \sum_{O\,=\,\O(\d^2)} b_O(M) \< \d^{(2)}(\vk)[O](\vk)\>' \vs
& + \sum_{O\,=\,\O(\d^3)} b_O(M) \< \d^{(1)}(\vk)[O](\vk)\>' \,,
\label{eq:Pk1h1l}
\ea
to simulations by rescaling it with a function $\upsilon(k,M)$.  Then,
by rescaling the profiles $y(k,M) \to y(k,M) \upsilon(k,M)$, the 
matter propagator $P_{1m}(k)$ should be described correctly to the extent
that the basic assumption of the halo model is valid.  A similar matching
of $P_{hh}(k,M,M')$ can be used to determine $P^{\eps\eps}_{MM'}(k)$.  

We leave this to future work, but point out that modifications to the
profiles $y(k,M)$ as a way to fit intermediate scales 
have also been proposed in \cite{mohammed/seljak}, who expanded the profile 
on large scales in powers of $k^2$.  
Ref.~\cite{baldauf/schaan/zaldarriaga} discussed a subtraction of the high-$k$ contribution of the loop integrals to \refeq{P1l} as an effective profile.  

Common to all these attempts at solving the issue of intermediate scales is the fact that we need to
introduce another scale that is larger than the
typical halo size (and close to the nonlinear scale).  This is clear from \reffig{Pk1l}, and given that
the typical wavenumber corresponding to halo radii is of order 
$k_{\rm HM} \sim \pi/R_{\rm HM} \sim 8 \iMpch$.  
The necessity of adding non-perturbative terms involving a new scale not directly related to halo profiles breaks with the
philosophy of the halo model as outlined in the introduction, i.e. that
predictions should be given completely in terms of perturbation theory
and well-defined properties of halos.  Moreover, the distinction between
deterministic and stochastic contributions is blurred in this
transition regime.  Nevertheless, the goal is to sufficiently constrain the additional terms to keep the halo model
predictive, in particular by considering various statistics such as
matter and halo power spectra and bispectra.   
We leave the whole question of intermediate scales as an open issue for future work.

\section{Bispectrum}
\label{sec:Bk}

We now consider the bispectrum (three-point function) of matter, and present the EHM prediction at tree level.  Given our findings
from \refsec{Pk}, we will drop the terms coming from the concentration
expansion, simplifying the expressions considerably.  At tree level, we then
need to expand \refeq{bias1} to second order.  This yields
\ba
\d_{h,M}(\vx,\tau) =\:& b_1(M,\tau) \d(\vx,\tau) + \frac12 b_2(M,\tau) [\d^2](\vx,\tau) \vs
& + \frac12 b_{s^2}(M,\tau) [(s_{ij})^2](\vx,\tau) \vs
& + [\eps](M,\vx,\tau) + [\eps_\d \d](M,\vx,\tau)\,,
\label{eq:dh2nd}
\ea
where we have slightly changed notation to match standard convention
for the density biases, and $s_{ij} \equiv (3\Om\cH^2/2)^{-1} (\partial_i\partial_j - \d_{ij}\nabla^2) \Phi$ is a scaled version of the tidal field.  The resulting expression
for the matter bispectrum is given in \refeq{BmApp}.  

The two main differences to commonly used halo model bispectra 
are that first, we are including the two second order bias terms, with respect
to density squared and tidal field squared [third and fourth lines in \refeq{BmApp}].  The \emph{tidal bias} \cite{chan/scoccimarro/sheth:2012,baldauf/etal:2012} has not been included in halo model calculations
of the bispectrum so far (e.g., \cite{smith/etal:07,Lazanu/etal:15}), but is straightforward to include once a parametrization of $b_{s^2}(M)$ is given which satisfies \refeq{conscond}.   

Second, the general stochastic terms which need to be included at this order
are given in the last line of \refeq{BmApp}.  Let us repeat them here:
\ba
& B_{mmm}(k_1,k_2,k_3) \stackrel{\rm stoch.}{=} \int d\rho(M_1)\int d\rho(M_2)\int d\rho(M_3) 
\vs
& \times  y(k_1,M_1) y(k_2,M_2) y(k_3,M_3) \,\Bigg\{   B^{\eps}_{M_1 M_2 M_3}(k_1, k_2, k_3) 
\vs
%
& \hspace*{2cm} + \left[ P_{\rm L}(k_2) P_{M_1 M_3}^{\eps_\d \eps}(k_3)
+ P_{\rm L}(k_3) P_{M_1 M_2}^{\eps_\d \eps}(k_2) \right] 
\vs
& \hspace*{2cm} + 2\  {\rm perm.}
%
\Bigg\} \,. \label{eq:Bmstoch}
\ea
where
\ba
B^{\eps}_{M_1 M_2 M_3}(k_1, k_2, k_3) \equiv\:& \Big\< [\eps](M_1,k_1) [\eps](M_2,k_2) [\eps](M_3,k_3) \Big\>'
\vs[4pt]
P_{M_1 M_2}^{\eps_\d \eps}(k) \equiv\:& \< [\eps_\d](M_1,\vk) [\eps](M_2,\vk') \>'
\,.  \label{eq:Bepsdef}
\ea
We thus need a parametrization of $B^{\eps\eps\eps}_{M_1M_2M_3}$ and $P^{\eps_\d \eps}_{M_1 M_2}$ (since we
are working at tree level, we do not need to perform any renormalization on $\eps$ and $\eps_\d \d$).  
In the high-$k$ limit, we can use the prediction of Poisson sampling from the local
deterministic halo abundance, \refeq{epsP3}.  At low $k$, \refeq{conscond} requires
that the mass-weighted integral of these quantities over any of the masses $M_i$ vanishes.  
We can immediately generalize our interpolating ansatz \refeq{Peps1} to $P^{\eps_\d \eps}_{M_1M_2}$ through
\ba
P^{\eps_\d \eps}_{M_1M_2}(k) =\:& \frac12 \frac{b_1(M_1)}{\bar n(M_1)} \d_D(\ln M_1-\ln M_2) 
\label{eq:Pepsd}\\
& - \frac12 \Theta_{M_1M_2}(k) \frac{M_1 M_2}{\rhob \< b_1 M \>_\rho} b_1(M_1) b_1(M_2)\,,
\nonumber
\ea
where
\be
\< b_1 M \>_\rho \equiv \int d\rho(M)\, b_1(M) M\,.
\ee
In \refapp{bispectrum} we also give a somewhat more lengthy expression for $B^{\eps\eps\eps}$ which satisfies the
corresponding constraints [\refeq{Beps}].  While it is a simple extension of \refeq{Peps1}, 
it clearly is not the only possible choice.  Again, we stress that further numerical investigations
of halo stochasticity, including its three-point function, as a function of scale are
necessary in order to obtain an accurate halo model bispectrum.  
 
Nevertheless, via \refeq{Pepsd} and \refeq{Beps}, and given bias parameters 
$b_2(M),\,b_{s^2}(M)$,  \refeq{BmApp} yields a consistent matter bispectrum obeying all
symmetries of the matter density, and asymptoting to the tree-level matter bispectrum
on large scales.  It does not include the effect of a modulation of halo profiles by
large-scale density perturbations, as we have found it to be numerically small in case
of the power spectrum, but this can be easily added back in.  Of course,
we expect the same issues on intermediate scales to arise that appear
for the power spectrum.   

\section{Matter velocity field}
\label{sec:vel}

So far, we have only considered the matter density field, since this is 
phenomenologically the most important quantity for large-scale structure.  
Let us now consider how the nonlinear matter velocities are described
in the self-consistent halo model approach pursued here.  First of all,
since the single-stream fluid picture breaks down on nonlinear scales,
our goal has to be to derive the velocity \emph{distribution} at a given
point $(\vx,\tau)$.  

Let us denote the matter velocity predicted by perturbation theory,
at the relevant order used in the halo model, by $\v{v}_{\rm PT}$.  
As argued in Ref.~\cite{MSZ}, halo velocities are unbiased with respect
to matter velocities up to higher derivative terms; that is,
the velocity of the effective halo fluid obtained by coarse-graining
the halo distribution is given by
\be
\v{v}_{h,M}(\vx,\tau) = \v{v}_{\rm PT}(\vx,\tau) + \O\left(
\nabla^2 \v{v}_{\rm PT},\, \bm{\nabla} \d_{\rm PT}\right)\,.
\ee
Since these higher derivative terms are assumed to be given entirely by the halo profiles 
in our approach, we set $\v{v}_{h,M} = \v{v}_{\rm PT}$.  
Then, the velocity distribution at $(\vx,\tau)$ is in the halo model
given by\footnote{Here, we are working in the non-relativistic limit and ignore corrections of order $\v{v}^2$.}
\ba
P(\v{v};\vx) =\:& \int dn(M) \int d^3 \vy \left[1 + \d_{h,M}(\vy)\right] \vs
&\quad \times
P_{v,h}(\v{v}-\v{v}_{\rm PT}(\vx); M, \vx-\vy)\,,
\label{eq:Pvx}
\ea
where we have dropped the time argument for clarity, and 
$\int dn(M) \equiv (\int \bar n(M) d\ln M)^{-1} \int \bar n d\ln M$
is the normalized integral over the halo number density (that is, without
mass weighting).  $P_{v,h}(\v{v}; M, \vr)$ denotes the mean normalized velocity
distribution within halos of mass $M$ at radius $\vr$.  By construction,
this obeys
\ba
\int d^3\v{v}\, P_{v,h}(\v{v}; M, \vr) =\:& 1 \quad\mbox{and} \vs
\int d^3\v{v}\, \v{v}\,  P_{v,h}(\v{v}; M, \vr) =\:& 0\,. 
\ea
For spherically symmetric halos, $P_{v,h}$ can be written as
\be
P_{v,h}(\v{v}; M, \vr) = P_v\left(v_\parallel = \v{v}\cdot\hat{\vr};\  
v_\perp = |\v{v} - (\v{v}\cdot\hat{\vr})\hat{\vr}|; M, r \right)\,,
\nonumber
\ee
i.e. in terms of the joint distribution of radial and tangential velocities.  
See \cite{smith/etal:08,lam/etal:13} for examples of modeling this 
velocity distribution.  

\refeq{Pvx} can then be generalized by allowing for a dependence of 
the velocity distribution on the halo concentration, for example, leading to
an expansion analogous to that discussed in \refsec{prof}.  
Further, one can straightforwardly apply the same type of reasoning
to obtain the momentum density, or mass-weighted velocity.

\section{Connection to the EFT}
\label{sec:EFT}

The EHM approach described in \refsec{HM} predicts, by construction,
a matter density field which matches the results of
perturbation theory to any desired order on large scales.  
Beyond the large-scale limit, the halo profiles lead to higher derivative 
terms $\propto \nabla^2 \d$, $(\nabla\d)^2$, and so on.  Further, the
halo model contains a stochastic contribution to the matter density field,
i.e. a contribution which does not correlate with long-wavelength perturbations.  
All these contributions satisfy the requirement of large-scale mass and momentum conservation
as long as the conditions \refeq{conscond} are satisfied.  

In this sense, this halo model approach consistently extends the 
predictions of the effective field theory (EFT) of LSS to nonlinear scales,
which necessarily implies that
the halo model is not guaranteed to be within a well-defined theoretical uncertainty
from the true answer when going beyond perturbative scales $k/k_{\rm NL} \ll 1$.  
A detailed study of the connections of the halo model to the EFT, while interesting,
is beyond the scope of this paper.  We will just make two comments of general interest here.

\textit{Matching to EFT parameters:}  as emphasized in \refsec{HM}, one key virtue of 
EHM is that it can be taken beyond tree level.  In the form that we
have defined the implementation there, corrections to the perfect
fluid description, i.e. the terms added by the EFT, are exclusively
provided by the halo profiles.  The results in \refsec{Pk1l} already
show however that the EHM ansatz fails to even roughly predict parameters such as the 
effective sound speed $c_s$:  EHM predicts a scale $k_{\rm HM} = \pi/R_{\rm HM}$,
while simulation measurements (e.g. \cite{Carrasco/etal:14}) find that
the correct scale is $k_{\rm NL} \ll k_{\rm HM}$.  There is no reason to
expect that this problem will be solved by higher loops; instead one has
to separately model the transition regime as discussed at the end of
\refsec{Pk1l}.

\textit{Higher derivative terms:} 
in the halo model, $\d_m$ is 
written as a convolution of a scalar $\d_{h,M}$ with a profile $y(r|M)$, 
where the profile is assumed to generate all higher derivative terms.  For this reason,
we only obtain higher derivative terms of the type $\partial^2 O$ and
$\partial_i O \partial^i O'$, where $O,O'$ are scalar operators appearing
in the expansions \refeq{bias1} and \refeq{yexp}.  
The second type is generated by having both $\partial^2 (OO')$ and $O (\partial^2 O')$ in the expansion.  This also holds when including the dependence of halo profiles and triaxiality 
on long-wavelength perturbations.  
Hence, the halo model, as described in \refsec{HM}, does not generate higher derivative terms of the form
$\partial_i s_{jk} \partial^k s^{ij}$, and similar terms for other non-scalar operators,
which are in general present in the EFT.  
This is a prediction which can be tested on simulations, by comparing the measured
amplitude (on scales within the perturbative regime) of higher derivative terms of the 
type $\partial_i s_{jk} \partial^k s^{ij}$ with, for example, $\partial_i s_{jk} \partial^i s^{jk}$.    
The halo model as described here only produces the second term.  
Of course, it is always possible to explicitly include any higher derivative term
in the expansion of the halo overdensity \refeq{bias1}.

\section{Conclusions}
\label{sec:concl}

We have presented a general procedure (\emph{EHM}) for constructing a halo model
description of the nonlinear large-scale structure which guarantees
mass and momentum conservation on large scales.  This procedure allows for
perturbation theory results to be matched to any given order.  Finally,
a single set of input ingredients (mass function, bias parameters, profiles,
and stochasticity covariances) describes all matter and halo auto- and
cross-correlations.  

We have attempted to write down the most general expression for the
matter density field that follows from the basic halo model assumption
stated at the beginning of \refsec{intro}, and that remains predictive.  
For this reason, we have only allowed terms at lowest order in derivatives
in the halo density expansion \refeq{bias1}, thereby declaring
that halo profiles are responsible for all higher derivative terms in the 
matter density field.  
While the number of input parameters in the model increases as one goes to higher
order (in particular the bias parameters of halos), these parameters 
can be measured in simulations (e.g., \cite{mcdonald/roy:2009,chan/scoccimarro/sheth:2012,saito/etal:14,lazeyras/etal}, or predicted via the peak-background
split approach for example.  The key virtues of the halo model, namely
simple, numerically cheap predictions for nonlinear matter statistics
on all scales that are physically motivated, are retained in any case.  

The new ingredients discussed here for the first time are the \emph{halo stochasticity 
covariance}, the \emph{concentration expansion} allowing for the
dependence of halo profiles on the environment, and a clarification
of the \emph{impact of low-mass halos} on halo model predictions.  
The last point, discussed in detail in \refapp{lowmass}, 
also applies to existing formulations of the halo model.  

Perhaps the most important conclusion of this work is that the
halo stochasticity covariance is a key ingredient of the halo model,
and likely to be numerically important in the transition region between
the classic 2-halo and 1-halo regimes.  This quantity 
has clearly not been studied in sufficient detail so far, 
with the most detailed studies being Ref.~\cite{hamaus/etal:2010,cai/etal:11}.  
Here, we have described a general procedure to derive the high-$k$ limit
of the stochasticity (in the 1-halo regime) in terms of the perturbative
bias parameters and mass function.    

The prescription given here does not by itself address the failure of the halo
model to describe the transition region between PT scales and the
1-halo regime.  In fact, going to 1-loop order in the power spectrum, we found that
the halo model performs worse than perturbation theory on scales
$k\sim 0.2-1\iMpch$.  This will most likely require additional ingredients
(see e.g. \cite{mohammed/seljak,baldauf/schaan/zaldarriaga} for related approaches).  
We leave this as a major open problem for future work.  

Turning to halo profiles, we have allowed for
the spherically averaged halo profiles as well as halo triaxiality to
depend on long-wavelength perturbations.  In order to avoid many free functions
of scale, we have parametrized this dependence only through the concentration.  
This however can easily be augmented to include the environmental dependence of halo outskirts as well.  Interestingly, we found that
halo triaxiality is likely to be unimportant in practice, as it is 
largely degenerate with the expansion of the spherically averaged profiles.  

A detailed comparison of the halo model power spectrum and bispectrum with simulation results is left for future work.  
This will also necessitate more study of the halo stochasticity.  
In order to be a fair comparison, this has to make use of state-of-the-art
numerically calibrated halo mass function, biases, and profiles.  

Let us also briefly discuss assembly bias, i.e. the fact that
the large-scale properties of halos depend on more than just the 
halo mass (e.g., \cite{gao/etal,wechsler/etal,dalal/etal}).  
In principle, assembly bias can be straightforwardly included
in the halo model, by promoting the integral over mass in \refeq{dmHM} 
to a multidimensional integral over mass, formation time, and/or other quantities.  
Correspondingly, the mass function $\bar n$, mean concentration $\bar c$, bias parameters $b_O$ and $b^c_O$, 
as well as stochastic fields $\eps_O,\,\eps_O^c$ all become functions of mass, formation time, and so on.  
Note that assembly bias can only affect halo model predictions if 
both profiles and biases and/or stochastic fields depend on additional
variables, for example, if at fixed mass halos with higher concentration
are more biased.  These effects thus only become relevant in the
intermediate to 1-halo regime.

Finally, the halo model can also be generalized to a model for galaxy
statistics via the halo occupation distribution (HOD) approach.  In the
spirit of the approach described here, the HOD for halos of a given mass 
should also be allowed to depend on the long-wavelength perturbations
via an expansion of the type \refeq{bias1}.  Of course, assembly bias effects
can also be included as described just above. 
These are expected to be more important for galaxy clustering than
for the matter density field; for example, certain types of galaxies may
live preferentially in early- or late-forming halos.  
We leave this to future work.    

\acknowledgments

I would like to thank Tobias~Baldauf, Mehrdad~Mirbabayi, Emmanuel~Schaan, Uro$\check{\rm s}$~Seljak, 
Masahiro~Takada, and Matias~Zaldarriaga for many helpful comments and discussions.  I gratefully acknowledge support from the Marie Curie Career Integration Grant  (FP7-PEOPLE-2013-CIG) ``FundPhysicsAndLSS.''

\appendix

\section{On the low-mass cutoff of halos}
\label{app:lowmass}

The halo model is based on parametrizations of the abundance, bias parameters,
and profiles of halos, all as a function of mass.  Clearly, these are only
calibrated over a certain mass range in simulations.  At high masses, 
there is no obstacle in principle to measuring halo properties accurately.  
A practical issue is that halos become exponentially rare at very
high masses.  However, this also makes them phenomenologically unimportant.  
For this reason, any extrapolation used at high masses is likely to be
well under control.  

On the other hand, properties of low-mass halos are poorly constrained by simulations due to
resolution limits.  The mass-weighting integrals, for example in 
\refeq{conscond}, converge very slowly towards low masses.  This raises
the question of whether the halo model predictions actually rely
on extremely low-mass halos whose properties are poorly known?

Here, we will show that this is not the case.  Consider the case where
the mass function, bias and profiles are well calibrated to a minimum
mass $M_s$.  We will show that the uncertainties to the halo model
predictions introduced by halos of mass below $M_s$ are of order 
$(k R_{M_s})^2$.  If the scales of interest are $k \ll 1/R_{M_s}$, then
this is a negligible uncertainty on the halo model predictions.  

To prove this, 
we introduce a low-mass cut-off $M_s$ so that all mass-weighting
integrals become
\be
\int d\rho(M) \to \int_{\ln M_s}^{\infty} \!\!\! d\ln M \frac{M}{\rhob} \bar n(\ln M)\,.
\ee  
In order to fix global mass conservation, we add an effective term
to the mass function at the cutoff,
\be
\bar n(M) \to \bar n(M) + \bar n_s \d_D(\ln M - \ln M_s)\,,
\label{eq:nbars}
\ee
where $\bar n_s$ is determined by requiring
\be
\intMs d\ln M \frac{M}{\rhob} \bar n(\ln M) + \frac{M_s}{\rhob} \bar n_s = 1\,.
\ee
Similarly, in order to ensure the consistency condition for $b_1$ [\refeq{conscond}], we
let
\be
b_1(M) = \left\{ \begin{array}{cc}
b_1(M), & M > M_s \\[3pt]
b_{1s}, & M = M_s \,,
\end{array}\right.
\label{eq:b1s}
\ee
and require
\be
\intMs d\ln M \frac{M}{\rhob} \bar n(M) b_1(M) + \frac{M_s}{\rhob} \bar n_s b_{1s} = 1\,.
\ee
Corresponding conditions are to be placed on the other biases $b_O(M)$.  
For simplicity, we restrict to the matter statistics in the linear version
of EHM here.  The deterministic contributions to $P_{1m}(k)$ and $P_{mm}(k)$
involve the following integral:
\ba
& \int d\rho(M) b_1(M) y(k,M) \to \vs
& \intMs d\ln M \frac{M}{\rhob} \bar n(M) b_1(M) y(k,M) + \frac{M_s}{\rhob} \bar n_s b_{1s} y(k, M_s) \vs[3pt]
& \stackrel{k R_{M_s} \ll 1}{=} 1 -k^2 \bigg[ \intMs d\ln M \frac{M}{\rhob} \bar n(M) b_1(M) a_M R_M^2 \vs
& \hspace*{2.4cm} + \frac{M_s}{\rhob} \bar n_s b_{1s} a_{M_s} R_{M_s}^2 \bigg]\,,
\ea
where in the last line we have used \refeq{b1s} and the low-$k$ limit
of the profile \refeq{ylowk}.  Taking the derivative
with respect to $\ln M_s$ of this expression, it easy to verify that
this result is independent of $M_s$ up to corrections of order 
$(k R_{M_s})^2$.  

We now consider the stochastic contribution.  In order to satisfy
\refeq{conscond}, we add an
additional stochastic field $\eps_s$ which only contributes to the
stochasticity of halos of mass $M_s$.  We require $\eps_s(\vk)$ to satisfy,
in the same sense as \refeq{conscond},
\be
\intMs d\ln M \frac{M}{\rhob} \bar n(M) \eps(M,\vk) 
+ \frac{M_s}{\rhob} \bar n_s \eps_s(\vk) = \O( R_{M_s}^2 k^2)\,.
\label{eq:epss}
\ee
Since $\eps_s$ is supposed to describe halos of mass $\leq M_s$, 
we require the scaling in terms of $R_{M_s}$ given on the r.h.s. of
\refeq{epss}.  Taking the auto-correlation of this equation then implies
\ba
&\Bigg\langle \Bigg[
\intMs d\ln M \frac{M}{\rhob} \bar n(M) \eps(M,\vk) + \frac{M_s}{\rhob} \bar n_s \eps_s(\vk) \Bigg]  \vs
&\times
\Bigg[
\intMs d\ln M' \frac{M'}{\rhob} \bar n(M') \eps(M',\vk') + \frac{M_s}{\rhob} \bar n_s \eps_s(\vk') \Bigg] \Bigg\rangle^\prime \vs[3pt]
& \propto (R_{M_s} k)^4 \,.
\ea
The stochastic contribution to $P_{mm}(k)$,
\be
\int d\rho(M) \int d\rho(M') P_\eps^{MM'}(k) y(k,M) y(k, M') \,,
\ee
can then easily be shown, via \refeq{ylowk} and \refeq{epss},  
to scale as $k^4$ and depend on $M_s$ only through terms of order $(R_{M_s} k)^2$. 

We conclude that, for $k < 1/R_{M_s}$ where $M_s$ is the lowest mass for
which halo properties are well calibrated, the halo model predictions
are under accurate theoretical control.

\section{Halo triaxiality}
\label{app:triax}

Dark matter halos are triaxial, and the orientation of the axes, as well
as having a random component, correlates with large-scale 
tidal fields $s_{ij}$.  
At linear order in the tidal field, this coupling 
can generally be of the form, dictated by symmetry,
\be
y(\vr,M,\tau,c)\Big|_{s_{ij}} = \left[1 + f_s^c(r, M,\tau)\,s_{ij} \frac{r^i r^j}{r^2}\right] y(r,M,\tau,c)\,,
\ee
where $f_s^c(r,M,\tau)$ is a general function.  Of course, in order
to retain the predictivity of the halo model, we would like to reduce
this to a number in analogy to the concentration expansion introduced above.  One possible choice
would be to assume that the tidal field distorts halos in a homologous 
way, 
\be
y(\vr,M,\tau,c)\Big|_{s_{ij}} = y\left(\sqrt{r^2 + b_s^c(M,\tau) s_{ij} r^i r^j},M,\tau,c \right)\,,
\ee
which implies $f_s^c = b_s^c (\partial\ln y/\partial\ln r)/2$.  However,
the integral over this quantity does not vanish, and thus violates the constraint \refeq{yconstraint}.  Let us thus instead choose, for illustrative purposes,
\ba
y(\vr,M,\tau,c)\Big|_{s_{ij}} =\:& y(r,M,\tau,c) 
\label{eq:ytidal}\\
& + b_s^c(M,\tau)\,s_{ij} \frac{\partial^i \partial^j}{\partial^2} y_c(r,M,\tau,c)\,, \nonumber
\ea
which satisfies \refeq{yconstraint}.  Note that for typical 
universal halo profiles (such as NFW or Einasto) the functions $y_c$
and $\partial y/\partial\ln r$ are very similar.  
\refeq{ytidal} is sufficient at linear order, but can be extended 
to higher order in the same way as \refeq{biasc}, including all terms
that have the same trace-free symmetric structure.  At quadratic order,
this will involve $\d\,s_{ij}$, $s_i^{\  k} s_{kj} - \d_{ij} (s_{kl})^2/3$
and $\eps_s^t\,s_{ij}$.

Now, when written in the form \refeq{ytidal}, the terms from $b_s^c$ and
higher order tidal coupling are all degenerate with terms in the concentration
bias expansion.  This is because higher derivatives are always contracted
with the $s_{ij}$, i.e. $\partial_i\partial_j s^{ij}$, 
$\partial_i s^{ik} \partial^j s_{jk}$ and so on, which brings them into the form
$\partial^2 \d,\, (\partial_i\d)^2$ etc.  However, what if we allow the tidal coupling 
to have a different $r$-dependence than the specific form $(\partial_i\partial_j/\partial^2) y_c$~?  At low $k$, the Fourier space version of $\partial y/\partial s_{ij}$, when enforcing mass conservation, has to be given by
\be
\frac{\partial y}{\partial s_{ij}}(k,M) \stackrel{k\to 0} = a R_M^2 k_i k_j + \O(k^4)\,,
\ee
where $a$ is a constant.  Again, this will lead to the same term at leading order as the concentration expansion.  

Finally, we should also take into account random triaxiality of halos;
this was the case studied by \cite{2006MNRAS.365..214S}.    
This can be achieved simply by replacing $s_{ij}$ in the relations above
with a stochastic trace-free tensor field $\eps^t_{ij}$.  The conclusions
remain the same: these terms scale in a very similar way as the 
stochastic terms in the concentration expansion.  

Thus, only by choosing a functional form for $\partial y/\partial s_{ij}$ that is significantly different from the profile expansion $\partial y/\partial\ln c$
can the tidal coupling of halo triaxiality produce a significant difference to the
terms already included in the concentration expansion.  Moreover, this difference will only
appear at fairly high $k$.  It thus seems likely that halo
triaxiality will be a subdominant component of the halo model.  
The terms found by
\cite{2006MNRAS.365..214S} in the halo model matter bispectrum would thus
be effectively captured, in our formulation, by second and third moments of $\eps^c$
and $\eps_\d^c$ (as well as their cross-correlations with $\eps_\d$ and $\eps$, respectively).  Note that we have not written these terms in \refsec{Bk}.

\begin{widetext}

\section{Halo model bispectrum}
\label{app:bispectrum}

Using the second order bias expansion \refeq{dh2nd}, and neglecting the concentration
expansion, we obtain the following result for the matter bispectrum in the halo model:
\ba
B_{mmm}(k_1,k_2,k_3) =\:& \int d\rho(M_1)\int d\rho(M_2)\int d\rho(M_3) \,  
 y(k_1,M_1) y(k_2,M_2) y(k_3,M_3) \,\Bigg\{   \label{eq:BmApp}\\
& b_1(M_1) b_1(M_2) b_1(M_3) B_{\rm T}(k_1,k_2,k_3) \vs[5pt]
& + b_2(M_1) b_1(M_2) b_1(M_3) P_{\rm L}(k_2) P_{\rm L}(k_3) 
+ 2\  {\rm perm.}
\vs
& + b_{s^2}(M_1) b_1(M_2) b_1(M_3) \left[\left(\hat{\vk}_2\cdot\hat{\vk}_3\right)^2 - \frac13\right] P_{\rm L}(k_2) P_{\rm L}(k_3)
+ 2\  {\rm perm.}
\vs
& + B^{\eps}_{M_1 M_2 M_3}(k_1, k_2, k_3)
+ \left[ P_{\rm L}(k_2) P_{M_1 M_3}^{\eps_\d \eps}(k_3)
+ P_{\rm L}(k_3) P_{M_1 M_2}^{\eps_\d \eps}(k_2) \right] + 2\  {\rm perm.}
\Bigg\}\,, \nonumber
\ea
where the tree-level matter bispectrum is given by
\be
B_{\rm T}(k_1,k_2,k_3) = 2 F_2(\vk_1,\vk_2) P_{\rm L}(k_1) P_{\rm L}(k_2) + 2\  {\rm perm.}\,,
\ee
and the stochastic terms are defined in \refeq{Bepsdef}.  

One possible form of $B^{\eps\eps\eps}$ that satisfies \refeq{conscond} in the low-$k$ limit, i.e. 
\be
\int d\rho(M_1) B^{\eps\eps\eps}_{M_1 M_2 M_3}(k_1, k_2, k_3) \stackrel{k_1 \to 0}{=} 0\,,
\ee
can be constructed as follows:
\ba
B^{\eps}_{M_1 M_2 M_3}(k_1, k_2, k_3)  =\:&
\frac{\d_D(\ln M_1 - \ln M_2)\d_D(\ln M_1 - \ln M_3)}{[\bar n(M_1)]^2} \vs
& + \Theta_{M_1 M_2 M_3} (k) \Bigg[- \frac{M_1 M_2}{\rhob \Mbar} \frac{\d_D(\ln M_2 - \ln M_3)}{\bar n(M_2)} 
+ \frac{M_1}{\rhob^2 \Mbar^2}
\left(M_1  -\frac13 \frac{\< M^2\>_\rho}{\Mbar} \right) M_2 M_3 \vs
& \hspace*{2.7cm} + 2\  {\rm cyclic~perm.} \Bigg]\,.
\label{eq:Beps}
\ea
Here, we have defined $\< M^2 \>_\rho \equiv \int d\rho(M) M^2$, and
generalized \refeq{finterp} to
\be
\Theta_{M_1M_2M_3}(k) = [1 + (k [R_{M_1}+R_{M_2}+R_{M_3}])^4 ]^{-1}\,.
\label{eq:finterp3}
\ee

\end{widetext}

\bibliography{REFS}

\end{document}